\documentclass[preprint,12p]{elsarticle}

\usepackage{amssymb, colortbl}

\journal{Nuclear Instruments and Method A}

\begin{document}
\begin{frontmatter}

\title{Vibrational excitation induced by electron beam and cosmic rays in normal and superconductive aluminum bars}

\author[label2,label3]{M.~Bassan}
\author[label1]{B.~Buonomo}
\author[label5]{G.~Cavallari}
\author[label2,label3]{E.~Coccia}
\author[label2]{S.~D'Antonio}
\author[label2,label3]{V.~Fafone}
\author[label1]{L.G.~Foggetta\fnref{fog}}
\author[label1]{C.~Ligi\corref{ligi}}
\author[label1]{A.~Marini}
\author[label1]{G.~Mazzitelli}
\author[label1]{G.~Modestino}
\author[label3,label1]{G.~Pizzella}
\author[label1]{L.~Quintieri}
\author[label1]{F.~Ronga}
\author[label4]{P.~Valente}
\author[label1]{S.M.~Vinko\fnref{vin}}

\address[label1] {Istituto Nazionale di Fisica Nucleare - Laboratori Nazionali di Frascati,\\ Via~E.~Fermi,~40~- 00044 Frascati, Italy}
\address[label2] {Istituto Nazionale di Fisica Nucleare - Sezione Roma2,\\ Via~della~Ricerca~Scientifica~- 00133 Rome, Italy}
\address[label3] {Dipartimento di Fisica, Universit\`a di Tor Vergata,\\ Via~della~Ricerca~Scientifica~- 00133 Rome, Italy}
\address[label4] {Istituto Nazionale di Fisica Nucleare - Sezione Roma1,\\ piazzale Aldo Moro 2~- 00185 Rome, Italy}
\address[label5] {CERN, CH1211, Gen\`eve, Switzerland}

\fntext[fog]{now at Laboratoire de l'Acc\`elerateur Linaire, CNRS-Orsay, 91898 Orsay cedex, France}
\fntext[vin]{now at Department of Physics, Clarendon Laboratory, University of Oxford, Parks Road, Oxford, OX1 3PU, UK}
\cortext[ligi]{Corresponding author}

\begin{abstract}

We report new measurements of the acoustic excitation of  an Al5056 superconductive bar  when hit by an electron beam,  in a previously unexplored temperature range, down to 0.35~K. 
These data, analyzed together with previous results of the RAP experiment obtained for $T$~$>$~0.54~K, show a vibrational response enhanced by a factor $\sim4.9$ with respect to that measured in the normal state. This  enhancement explains the anomalous large signals due to cosmic rays previously detected in the NAUTILUS gravitational wave detector.

\end{abstract}

\begin{keyword}Gravitational wave detectors \sep Cosmic rays \sep Radiation acoustics
\PACS 04.80.Nn \sep 41.75.Ht \sep 96.50.S- \sep 61.82.Bg \sep 74.70.Ad
\end{keyword}

\end{frontmatter}

\thispagestyle{plain}

\section{Introduction}

Cosmic ray showers can excite sudden mechanical vibrations in a metallic cylinder at its resonance frequencies; in experiments searching for gravitational waves ($gw$) these disturbances are hardly distinguishable from the searched signal and represent an undesired source of accidental events, thus increasing the background. This effect was suggested many years ago and a first search was carried out with limited sensitivity with room temperature Weber type resonant bar detectors and ended  with a null result~\cite{Ezrow:1970yg}. Later on, the cryogenic resonant $gw$ detector NAUTILUS~\cite{rog} was equipped with a streamer tube extensive air  shower detector~\cite{coccia} and the interaction of cosmic ray with the antenna has been studied in detail. This apparatus allowed the first detection  of cosmic ray signals in a $gw$ antenna, that took place in 1998, when NAUTILUS was operating at a temperature $T$~=~0.14~K~\cite{cosmico1},  i.e. below the superconducting ($s$) transition critical temperature $T_c \simeq$~0.9~K.  During this run many events of very large amplitude were detected. This unexpected result prompted the construction, in 2002, of a scintillator cosmic ray  detector  also for the EXPLORER $gw$ detector as well as the beginning of a dedicated experiment (RAP)~\cite{rap}, that was planned at the INFN Frascati National Laboratory to study the vibration amplitude of a small Al5056 bar caused by the hits  of a 510 MeV electron beam. The experiment was also motivated by the need of a better definition of the thermophysical parameters of the alloy Al5056, used in the bar detector, at low temperatures. A detailed study of this effect is indeed useful to study the performance of $gw$ bar detectors for exotic particles~\cite{Astone:1992zf} and  to understand the noise due to cosmic rays in interferometric $gw$ detectors~\cite{Yamamoto:2008fs}. In this paper we summarize our previous knowledge on this effect. We then report the final results of the RAP experiments presenting measurements down to 0.35 K, and show how these new data help in sheding light on the 1998 anomalous NAUTILUS high energy events. We also recall that the detailed study of this thermo-acoustic effect has applications in devices used to monitor and measure particle beams characteristic,  and in particular in monitoring high power beams~\cite{kali}.

\section{The Thermo-Acoustic Model}\label{TAM}

\subsection{The model: normal conductive $(n)$ state}
B.L.~Beron and R.~Hofstadter~\cite{beron1,beron2}  first measured mechanical oscillations in piezoelectric disks hit by a high energy electron beam. The authors first pointed out that cosmic ray events could excite mechanical vibrations in a $gw$ metallic antenna and that, consequently, cosmic rays could represent a background for experiments aimed at the detection of $gw$. The interaction of a ionizing particle with the bulk of a suspended cylindrical bar generates a pressure pulse in the bar. More in detail, the energy lost by the particle in the bar causes a local warming up of the material;  the local thermal expansion in the bulk generates the pressure wave. This sonic pulse determines the excitation of the vibrational elastic modes of the suspended bar.

A.M.~Grassi~Strini et al.~\cite{grassi} reported the results of an experiment based  on a pure aluminum bar exposed to a proton beam. The experimental data were compared to a theoretical  model based on the Fourier response of a thin bar to the pressure wave originated by a delta-like thermal perturbation. If the heating is in the bar center, only even Fourier harmonics are allowed.  The ``maximum  amplitude of oscillation for the fundamental longitudinal elastic mode'', in the following referred to as ``Amplitude", for a  material in a normal ($n$) state of conducting is given, according to this Thermo-Acoustic Model (TAM), by:
\begin{equation}
B_n^{~th}=\frac{2 \alpha L W}{\pi c_V M} 
\label{b0}
\end{equation}
\noindent where the suffix "$th$" stand for the theoretically expected value. This result applies to a thin cylinder (with radius $R$ and length $L$, $R \ll L$ and mass $M$), for a beam hitting on center of the cylinder lateral surface. Here $W$ is the total energy released by the beam to the bar,  $\alpha$ is the linear thermal expansion coefficient and  $c_V$ is the isochoric specific heat. The dimensionless Gr\"{u}neisen parameter $\gamma$ of the material includes the $\alpha/c_V$ ratio:
\begin{equation}
\gamma=\frac{\beta K_{T}}{\rho c_{V}}
\label{gru}
\end{equation}
\noindent where $\beta$ is the volume thermal expansion coefficient  ($\beta=3\alpha$ for cubic elements), $K_{T}$ is the isothermal bulk elastic modulus and $\rho$ is the material density. The Gr\"{u}neisen  $\gamma$ slightly depends on the temperature when the material is in  the $n$ state.

Eqn.~(\ref{b0}) is a limit case of a more general problem, when the paths of the interacting particles in the bulk~\cite{allega,deru,liu} are considered. Introducing a vector field $\mathbf{u}(\bold{x},t)$ describing the local displacements from equilibrium, the amplitude of the $k-th$ mode of the cylinder oscillation is proportional to:

\begin{eqnarray}
g{_k}{^{therm}}& = & {\frac{\Delta P^{therm}}{\rho}} {\cal A'} {\cal I}{_k} \nonumber \\
                          & =  & {\frac{\gamma}{\rho}}\left| {\frac{dW}{dx}}\right| {\cal I}{_k}
\label{gtherm}
\end{eqnarray}
\noindent where  $\Delta P^{therm}$ is the pressure pulse due to the sonic source described above, $dW/dx$ is the specific energy loss of the interacting particle, $\cal A'$ is the cross section of the tubular zone centered on the particle path in which the effects are generated and ${\cal I}{_k}=\int{dl (\nabla\cdot\bold{u}{_k}(\bold{x}))}$ is a line integral over the particle path involving the normal mode of oscillation $\bold{u}{_k}(\bold{x})$. The Amplitude predicted by Eqn.~(\ref{b0}), can be rederived from Eqn.~(\ref{gtherm}) in the simplified case of a thin bar $(R/L\ll 1$) and for a particle hitting on the bar center. We can therefore adjust the value of Amplitude predicted by Eqn.~(\ref{b0}) to a more correct value:
\begin{equation}
B_n^{~th}=\frac{2 \alpha L W}{\pi c_V M} (1+\epsilon) 
\label{xth}
\end{equation}
where $\epsilon$ is a corrective parameter estimated by a Monte Carlo (MC) simulation~\cite{rap}, which takes into account the solutions $O[(R/L)^2]$ for the modes of oscillation of a  cylinder, the transverse dimension of the beam at the impact point  and the trajectories of the secondary particles generated in the bar. The value of $\epsilon$ for the bar used in the experiment is estimated by MC to be --0.04.

\subsection{The model: superconducting $(s)$ state}
When the material is in the $s$ state,  a part of the energy lost by particle causes the suppression of the superconductivity in a region, called hot spot, that is centered around the particle path.
The maximum possible radius of the hot spot, $r_{HS}$, is obtained by equating the specific energy lost by the particle, $dW/dx$, to the enthalpy variation (per unit volume), $\Delta h$, for the transition from the $s$ state at temperature $T$ to the $n$ state~\cite{wood,gray,sherman,strehl}:
\begin{equation}
r_{HS} = \sqrt{\frac{\cal A ''}{\pi}} = \sqrt{\frac{|dW/dx|}{\pi\ \Delta h}}\
\label{rHS}
\end{equation}
\noindent  where $\cal A''$ is the cross section of the zone switched to the $n$ state. 

\noindent The creation of a hot spot by a particle interacting with a material in $s$ state causes a further correction to Amplitude of Eqn.~(\ref{xth}), a term which is peculiar to the particle propagating in a zone now switched to the $n$ state. The additional contribution to the amplitude of the cylinder oscillation mode $k$ is proportional to \cite{allega,deru}:
\begin{eqnarray*}
g{_k}{^{trans}}&=& {\frac{\Delta P^{trans}}{\rho}} {\mathcal A''} {\mathcal I}{_k}\\
       &=& {\frac{1}{\rho}}\left[{ K_{T}\frac{\Delta V}{V}+\gamma T \frac{\Delta \mathcal{S}}{V}}\right]  {\mathcal A''}{\mathcal I}{_k}
\end{eqnarray*}
\noindent Here $\Delta V$ and $\Delta \mathcal{S}$ are the differences of the volume and entropy in the two states of conduction. The differences can be expressed in terms of the thermodynamic critical field $H_c$ and it follows~\cite{hake}, in first approximation, that\footnote{We keep the  CGS electro-magnetic system of units  for the magnetic field, as used by the authors of the cited articles, and we convert the density of the magnetic energy to SI units.}:
\begin{eqnarray*}
\frac{\Delta V}{V} = \frac{V_n-V_s}{V} = \frac{H_c}{4 \pi} \frac{\partial H_c}{\partial P}
\end{eqnarray*}
and
\begin{eqnarray*}
\frac{\Delta \mathcal{S}}{V} = \frac{\mathcal{S}_{n} - \mathcal{S}_{s}}{V} = -\frac{H_c}{4\pi} \frac{\partial H_c}{\partial T}
\end{eqnarray*}
The quadratic dependence $H_c(t)=H_c(0)(1-t^2)$ on $t$, where $t=T/T_c$, is assumed in computing the differences.

\noindent Therefore the value of Amplitude due to a particle creating hot spots in a material in $s$ state is given by: 
\begin{eqnarray}
B_s^{~th}&=& B_n^{~th} \left( 1 + {\mathcal R}\right) \nonumber \\
 &=& B_n^{~th} \left[1 + \left( \Pi \frac{\Delta V}{V}+T \frac{\Delta \mathcal{S}}{V} \right) \left({ \Delta h}\right)^{-1} \right] \ 
\label{exp}
\end{eqnarray}

\noindent where ${\cal R} = g{_k}{^{trans}}/g{_k}{^{therm}}$ and the definition (\ref{gru}) of  $\gamma$ is used to obtain:

\begin{displaymath}
\Pi = \frac{2\rho L (1+\epsilon)}{3\pi M\frac{B_n^{~th}}{W}}
\end{displaymath}

\noindent  Eqn.~(\ref{xth}) and (\ref{exp}) show, by inspection,  that the Amplitude $B^{~th}$ linearly depends on $W$, the energy released by particle, both in the $n$ and in the $s$ states. 
Therefore it appears natural to consider, as we do in the following, the ratio $B^{~th}/W$ as a measure of the relevant material properties. Finally we note that the knowledge of the specific heat of the material for the $s$ state, $c_s$, allows us to approximate the exact Eqn.~(\ref{rHS}) with the following condition for the transition $s \to n$ of a volume $V$ of the material at temperature $T$,  due to the absorption of energy $W$ from the particle \cite{wood, sherman}:
\begin{equation}
W>VC{_I}(T)
\label{swisn}
\end{equation}
    
\noindent with:
\begin{displaymath}
C_I(T)=\int_T^{T_c}c_s(T') dT'
\end{displaymath}

\noindent Moreover, the knowledge of $c_s(T)$ allows to derive $\Delta h$ from the relation $\Delta h(T) = C_I (T) + T\frac{\Delta {\mathcal S}}{V}$ ~\cite{dwhite}.

\subsection{Amplitude predictions for the normal and superconductive state} 
In order to compute, by means of Eqn.~(\ref{xth}), the expected value of Amplitude at different temperatures, we need, for the $n$ state, both $\alpha(T)$ and $c_{V}(T)$ of the material. 
As these values are not well known for the Al5056 alloy, we used those of pure aluminum. Polynomial interpolations on data of Ref.~\cite{kroeg} (12~$<T\leq$~300~K) and the parametrization in Ref.~\cite{coll} ($T\leq$~12~K)  give $\alpha(T)$, while $c_{V}(T)$ is obtained by polynomial interpolations on values of $c_{P}$ reported in Ref.~\cite{crc}. Table~\ref{tb10}  shows the computed values of $\alpha$,  $c_{V}$ and the normalized Amplitude $B_n^{~th}/W$.
\begin{table}[t]
\centering
\begin{tabular}{|c|c|c|c|}
\hline
 $T$& $\alpha $ & $c_{V}$  & $B_n^{~th}/W $\\
 $[\rm{K}]$& $[10^{-6}$ K$^{-1}]$ & [J mol$^{-1}$ K$^{-1}]$  & $[10^{-10}$ m J$^{-1}]$ \\
\hline
264           &22.2               & 23.5                        &  2.23 \\
71             &7.5                 &7.94                         &  2.23 \\
4.5            &5.8~$\times$~10$^{-3} $ &7.6~$\times$~10$^{-3}$   & 1.80  \\
1.5            &1.5~$\times$~10$^{-3} $ &2.1~$\times$~10$^{-3}{^\dag}$ & 1.72 \\
\hline
\end{tabular}
\caption{\it Normal state of conduction. Amplitude normalized to the beam released energy W and input values for the calculation ($\alpha, c{_V}$) in the case of the RAP bar ({L~=~0.5~m; M~=~34.1~kg}) made of pure aluminum. The correction of Eqn.~(\ref{xth}) is applied. $^\dag c_V$ value for Al5056.}
\label{tb10}
\end{table}

For the $s$ state instead, measurements performed at  very low temperatures \cite{barucci2010} on samples belonging to the same production batch of our Al5056 bars, allow us to characterize the relevant properties of that alloy. The measurement of the transition temperature to the $s$ state, carried out using the mutual inductance method, yields  $T_c = 0.845 \pm 0.002$~K and a total transition width of  about $0.1\ \rm{K}$.  Few data of specific heat for Al5056 are available in the literature~\cite{coccianin}; however, new  $c_V$ measurements were performed above and below $T_c$ \cite{barucci2010}. 

In the temperature interval 0.9~$\leq$~T~$\leq$~1.5~K, i.e.~in the $n$ state, assuming for the specific heat the usual low temperature parametrization: $c{_V}$~=~$\Gamma T$~+~$\Psi T^{3}$, 
the measurements give the values $\Gamma$~=~1157~$\pm$~31~erg~cm$^{-3}$~K$^{-2}$ for the electronic specific heat coefficient and $\Psi$~=~140~$\pm$~10~erg~cm$^{-3}$~K$^{-4}$ for the lattice contribution. 

On the other hand, computing  the Amplitude in the $s$ state by means of Eqn.~(\ref{exp}) requires the knowledge of a) the thermophysical parameters $\alpha_n$ and $c_{V,n}$ of the material, in order to evaluate $B_n$ for the $n$ state below $T_c$ and b) the dependence of  $H_c$ on $T$ and $P$ for calculating the derivatives $\partial H_c/\partial T$ and $\partial H_c/\partial P$.
The requirement a) cannot be fulfilled due to the lack of knowledge of $\alpha_n$ for Al5056 and we therefore assume for $B_n/W$ the  value  measured just above $T_c$.  This assumption is justified by the fact that $\gamma_n$ usually has, below $T_c$, a very weak dependence on temperature. Regarding requirement b), we derive $\partial H_c/\partial T$ at $T<T_{c}$ from the $H_c$ parabolic dependence  on $t$; we also assume that the unknown dependence of $\partial H_c/\partial P$ on $t$ at $P=0$ for Al5056 is equal to that of pure aluminum and, therefore, can be obtained from the tabulation  of $H_c$ as a function of $T$ and $P$ contained in Ref.~\cite{harris}. If the superconducting properties of Al5056 can be described by the BCS theory, then $H_c(0) \approx 2.42\  \Gamma{^{1/2}} T_c \approx 70\ \rm{Oe}$. Insertion of the numerical values in the Eqn.~(\ref{exp}) yields to values of $B_s/W$ ranging from --9.2~$\times$~10$^{-10}$ to --7.3~$\times$~10$^{-10}$~m J$^{-1}$ in the temperature interval having limits 0.3 and 0.8~K, respectively. The lower limit of the temperature interval is constrained by the data availability in the  $H{_c}(P,T)$ tabulation  of Ref.~\cite{harris}.

\section {The RAP experimental setup} 

\subsection {The bar and the piezoelectric ceramics}
The  RAP experiment has been fully described in Ref.~\cite{rap}. Here we briefly recall that the test mass is a  cylindrical bar ($R$~=~0.091~m, $L$~=~0.5~m, $M$~=~34.1~kg) made of Al5056, the same aluminum alloy (nominal composition 5.2\% Mg and 0.1\% of both Cr and Mn) used for NAUTILUS. The bar hangs from the cryostat top by means of a multi-stage suspension system ensuring attenuation from the external mechanical noise of  --150 dB in the 1700--6500 Hz frequency window. The frequency of the fundamental longitudinal mode of oscillation of the bar is 
$f_0$~=~5414.31~Hz below $T$~=~4~K. 

At  temperatures below 10~K, the Al5056 intrinsic Q factor is 4.1~$\times$~10$^7$~\cite{coccia84} corresponding to a decay time of the order of 20 minutes:  a shorter decay time is desirable in order to have a more manageable repetition rate of the hits. This is simply achieved by the presence of a thermometer on one of the bar end face, that damps the oscillations to a decay time of the order of 30 seconds. Therefore, we only have to wait  a couple of minutes between consecutive hits to avoid the pile-up of the signals. 
 
Two piezoelectric ceramics (Pz), electrically connected in parallel, are inserted in a slot milled out in the center section of the bar, opposite to the bar suspension point, and are squeezed when the bar shrinks. In this Pz arrangement the strain measured at the bar center is proportional to the displacement of the bar end faces. The Pz output is first amplified,  and then sampled at 100~kHz by an ADC embedded in a VME system, hosting the data acquisition system. A band pass filter between 300~Hz and 50~kHz is used to reduce the low frequency power line noise and to avoid Fourier aliasing. 
  
The measurement of the Pz conversion factor $\lambda$, relating voltage to oscillation amplitude, is accomplished according to a procedure~\cite{calib} based on the injection in the Pz of a sinusoidal waveform of known amplitude, with frequency $f_0$ and time duration less than  the decay time of the mechanical excitations and on the subsequent measurement of Amplitude. The procedure is correct if $R/L \ll 1$ and carries a 6\% systematic error. 

The value of $\lambda$ during the 2009 run was 1.26~$\times$~10$^7$~V m$^{-1}$ at room temperature and  1.16~$\times$~10$^7$~V m$^{-1}$  constant at temperature $T\le$~4.5~K.

\subsection{The Frascati DA$\Phi$NE Beam Test Facility}

The DA$\Phi$NE BTF transfer line  can transport and deliver in controlled way  electron or positron  beams, from the end of the DA$\Phi$NE Linac  to a 100~m$^2$ experimental hall, where users normally set up and operate their tests and experiments. Particles can be provided  in 1 or 10~ns duration pulses, with an injection frequency\footnote{Actually one of the 50 pulses in 1 s is sent to an hodoscope in order to reconstruct the energy profile of the beam at the end of the Linac} spanning from 1 to 49~Hz, in a wide energy range (25~--~750~MeV for electrons) and intensity (from 1 up to $10^{10}$~particles/pulse).

Since the end of 2004, when the RAP experiment was first operated with the aluminum bar, some important upgrades \cite{BTF_upgrade} and diagnostic improvements \cite{BTF_diagnostic} have been accomplished. The most important improvement concerned the installation, during the 2006 shutdown, of a pulsed dipole magnet at the end of the Linac: such magnet allows to alternate the beam between the DA$\Phi$NE damping rings and the test beam area, thus enhancing the BTF duty cycle, due to the reduced switching time. So that, when beams are  injected into the collider, the  BTF  can still receive beam, although with a lower repetition rate. The continuous feeding of the BTF line, during the time sharing with the DA$\Phi$NE collider, is technically  possible because not all the Linac bunches are needed for  filling the  accumulator rings. 
Obviously, in this operation scheme, the pulse length and the primary beam energy are the same as those fed to the DA$\Phi$NE collider. Nevertheless  this does not constitute  a real limitation, since the facility is mainly operated in single particle mode (electrons/positrons), which is the ideal configuration for detector calibration and testing. In this case the beam characteristics are mainly determined by the  magnetic field of the energy selector (the upstream 45$^{\circ}$ dipole, on the BTF transfer line)  and by  scrapers suitably positioned along the line itself.

\begin{figure}[t]
\begin{center}
\includegraphics[width=9cm]{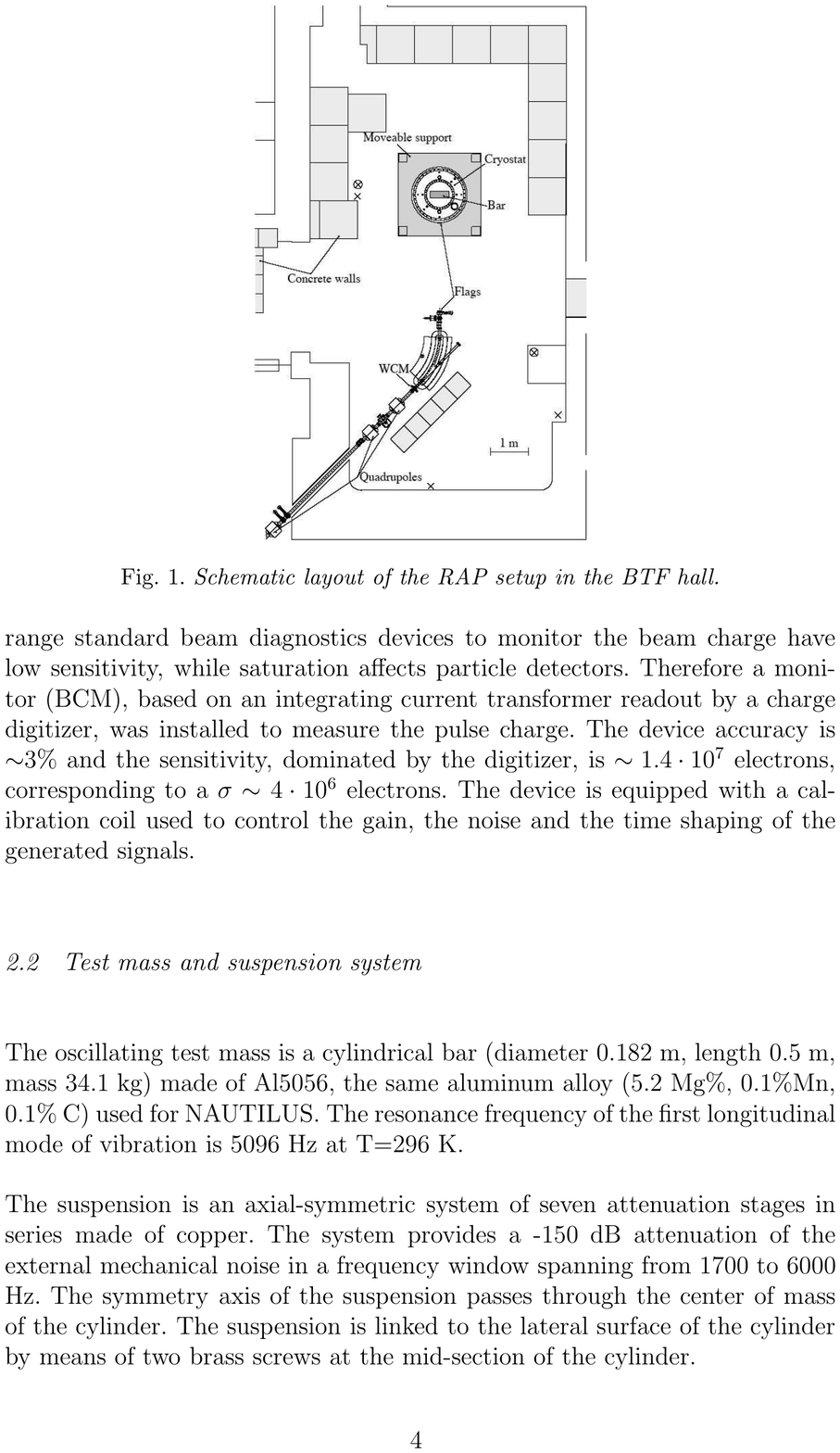}
\caption{\it Schematic layout of the RAP setup  in the  BTF hall.}
\label{set_up}
\end{center}
\end{figure}

The RAP experimental set-up was installed  about 2.5~m away from the exit of the beam (see Fig.~\ref{set_up}). The correct position of the beam at the exit of the line and at the entrance of the cryostat was monitored shot-by-shot by two high sensitivity fluorescence flags, 1~mm thick alumina doped with chromium targets. The spot size at the entrance of the cryostat, 50~cm far from the center of the bar, is about 2~cm in diameter. The Monte Carlo simulation of the detector indicates that the beam spot at the surface of the cylindrical bar preserves almost unchanged, due to  the negligible effects of the cryostat vacuum and thermal shields intercepted by the beam.

\begin{figure}
\begin{center}
\includegraphics[height=7.5cm,width=10cm]{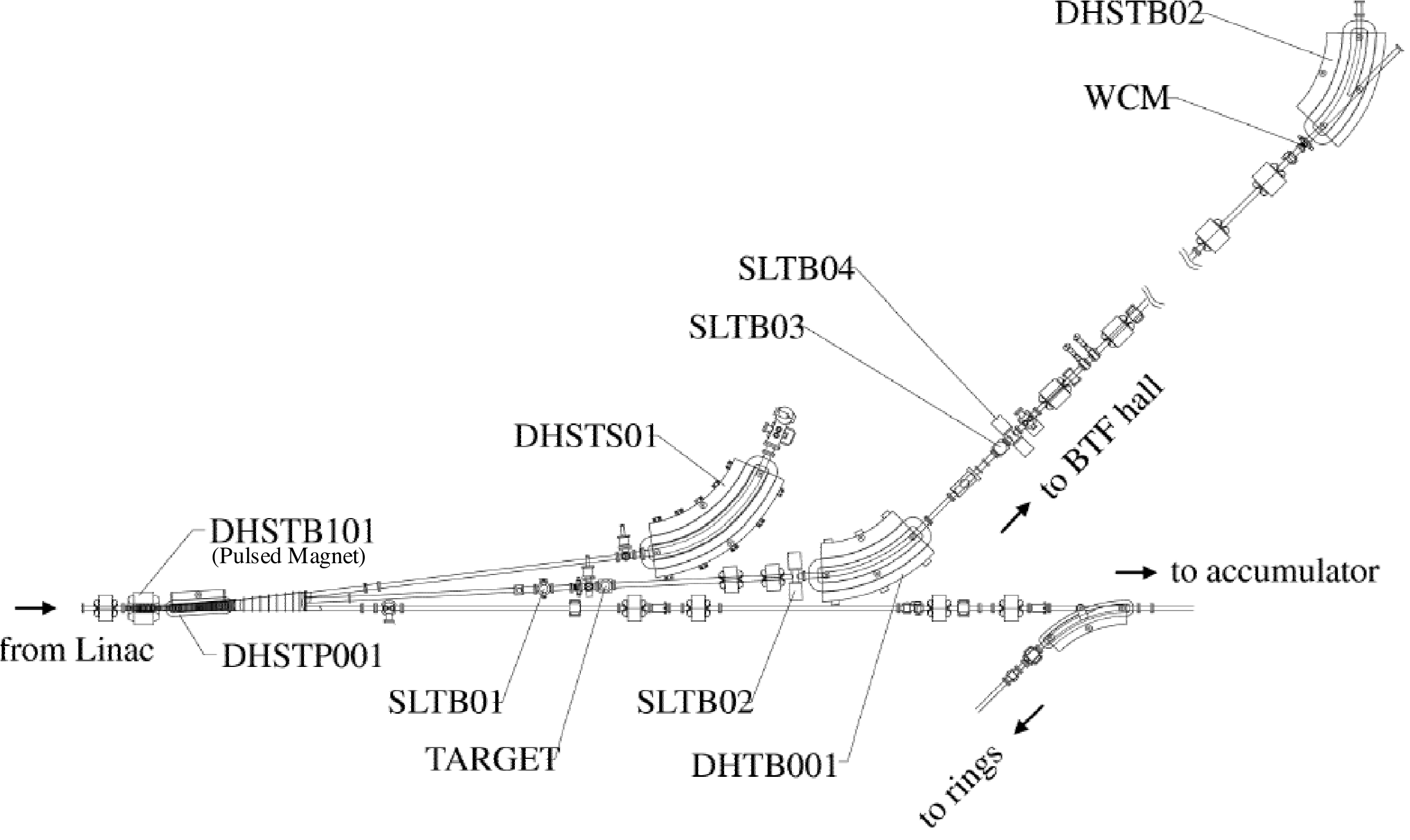}
\caption{\it Magnetic layout of the BTF transfer line: DHSTP001 is the pulsed dipole magnet added during 2006 upgrade.}
\label{layout} 
\end{center}
\end{figure}

The beam multiplicity for the RAP measurements performed in 2009 spanned from $10^{7}$  to $10^{9}$ electrons per bunch of 10~ns width. The current intensity modulation was obtained by properly changing the aperture of the  tungsten slits (both horizontal and vertical), along the BTF transfer line (SLTB\emph{ij} in Fig.~\ref{layout}): this procedure, unlike that used in the previous RAP measurement campaign and based on the defocusing of the beam by quadrupoles, allows us to avoid the beam degradation, since the particles in the external tails are cut away.
 
For the beam diagnostics, a monitor (WCM), based on an integrating current transformer, readout by a charge digitizer, was used to measure the pulse charge. The device accuracy is 3\% and the readout noise fluctuations give a measurement error $\sigma$~=~1.5~$\times$~10$^{7}$ electrons. The device is equipped with a calibration coil used to control the gain, the noise and the time shaping of the generated signals.

\subsection{Cryogenic setup}

\begin{figure}
\begin{center}
\includegraphics[height=17cm]{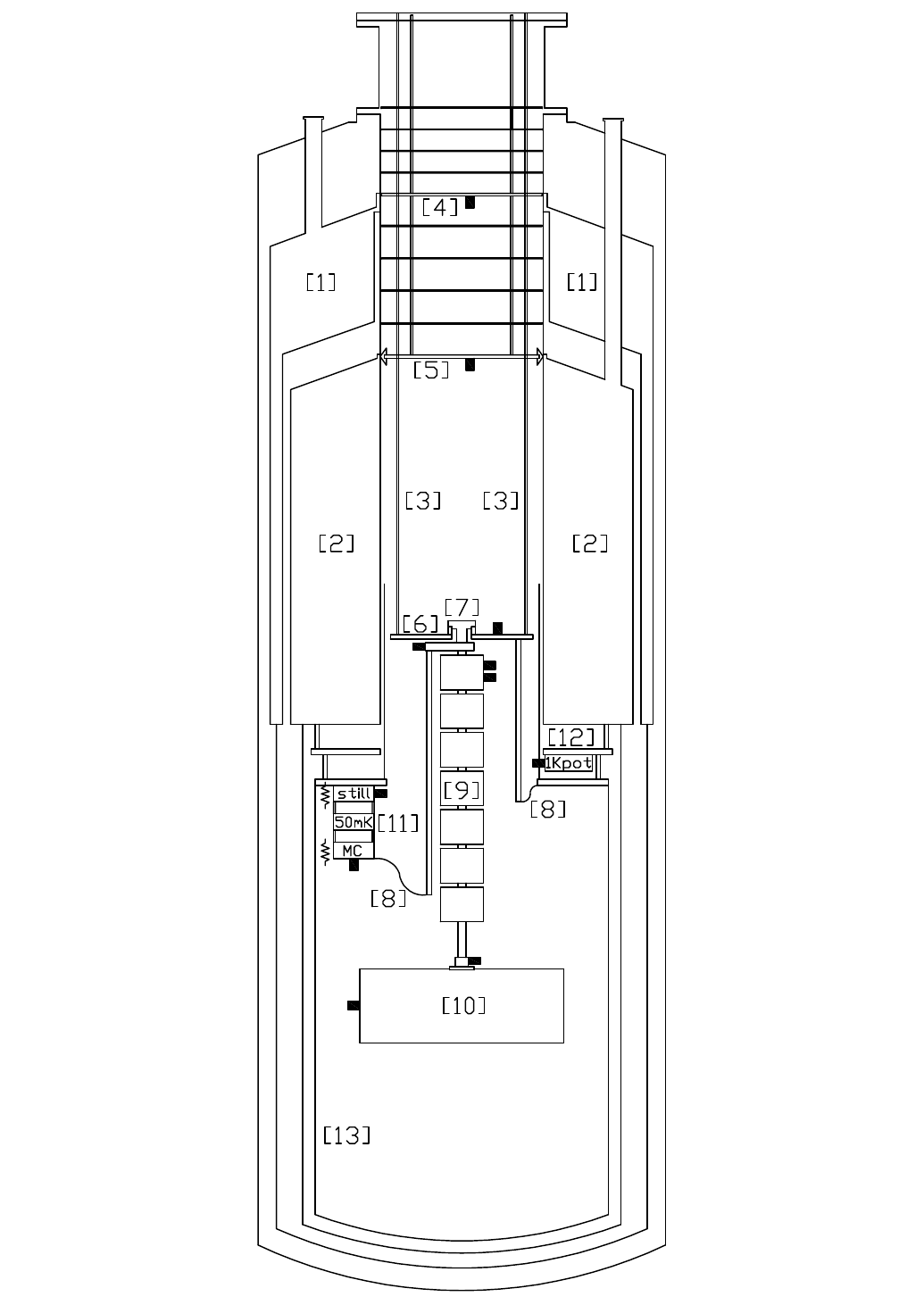}
\caption{\it Schematic view of the RAP cryostat: [1] LN$_2$ reservoir, [2] LHe container, [3] Stainless Steel (SS) suspension cables, [4] 77 K flange, [5] 4.2 K flange, [6] 0.6 K flange, [7] SS screw with Teflon ring, [8] soft copper thermal contacts, [9] copper suspension, [10] bar,  [11] dilution refrigerator cold end, [12] 1K Pot, [13] radiation shields. 
 The filled squares represent the thermometers.}
 \label{cryo}
 \end{center}
\end{figure}

The RAP cryogenic setup consists of a KADEL commercial liquid helium cryostat, 3.2 m high and 1 m in diameter, suspended on a vertically movable structure, and containing a dilution refrigerator. A schematic view of the cryostat together with the cold side of the dilution refrigerator is depicted in Fig.~\ref{cryo}. The liquid helium (LHe) and liquid nitrogen (LN$_2$) dewars, with a capacity of 340~L and 200~L respectively, are placed in the upper half. Three stainless steel cables are suspended from the top flange to support the experimental apparatus. To avoid the radiation input, 8 aluminum radiation shields are mounted between the top flange at room temperature, the 77~K OFHC (Oxygen Free, High Conductivity) copper flange and the 4.2~K OFHC copper flange. These two latter flanges are mechanically connected with the LN$_2$ and LHe dewars, respectively. The experimental chamber is positioned on the lower half of the cryostat and is surrounded by one OFHC copper radiation shield connected to the Still flange, two aluminum radiation shields connected to the LHe dewar and the LN$_2$ dewar respectively, and the outer aluminum container. The LHe container is indium sealed to separate the experimental chamber volume from the insulation zone.

\begin{figure}[t]
\begin{center}
\includegraphics[height=10cm,width=12cm]{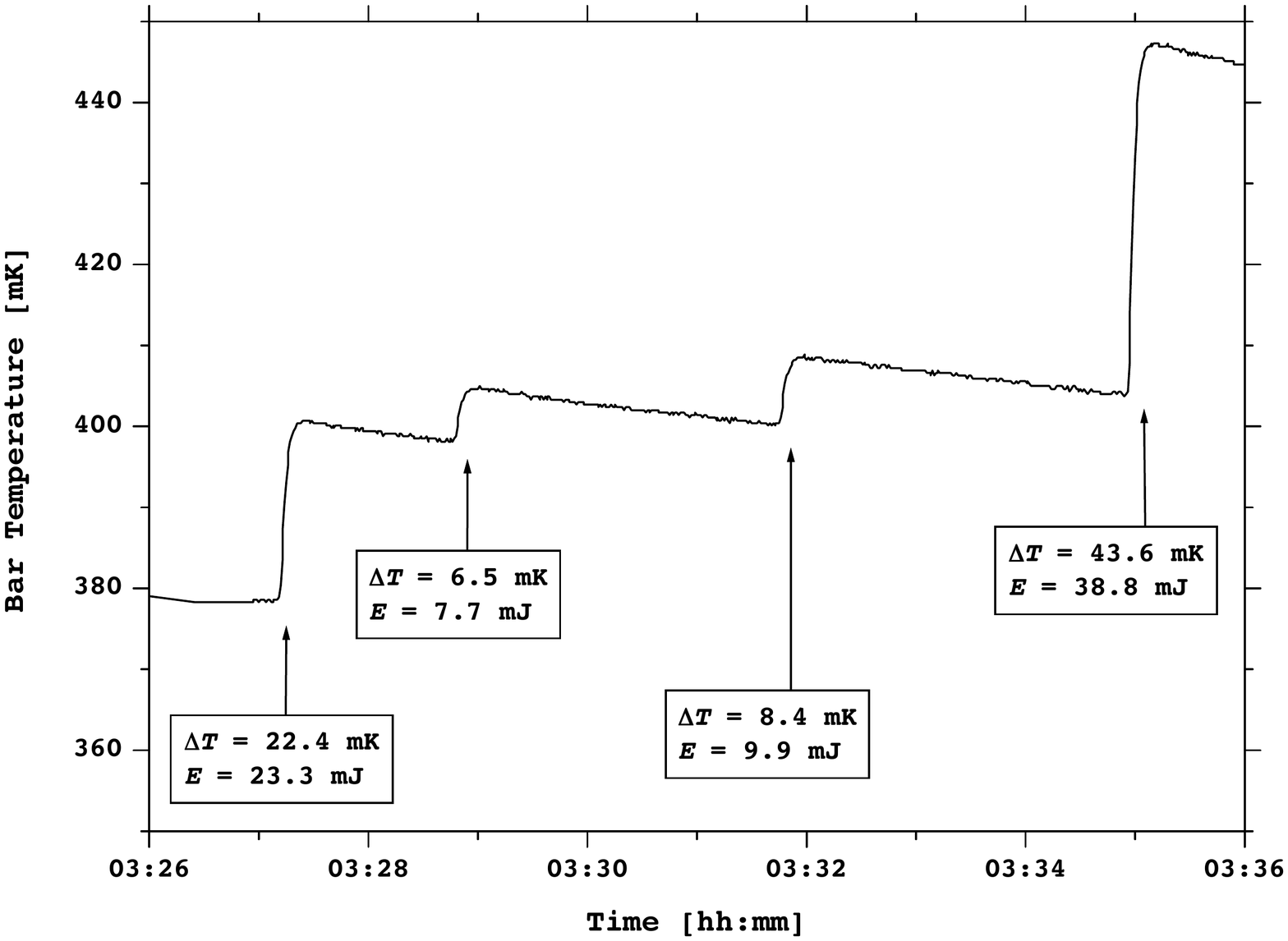}
\caption{\it Bar temperature measured during a run. Four beam shots are visible. The steps in temperature are proportional to the energy deposited in the bar. They can be used to measure the Al5056 specific heat, yielding a result in agreement with the measurement in Ref.~\cite{barucci2010}. Moreover, this information is useful to monitor the beam and to check the intensity measured by the current monitor.}
 \label{jump} 
 \end{center}
\end{figure}

The cryostat hosts a continuous flow, closed cycle $^3$He-$^4$He dilution refrigerator, made by Leiden Cryogenics, with a base temperature of about 100~mK and a cooling power of about 1~mW at 120~mK. The continuous flow is ensured by a pumping system, composed of two Varian TV~551~NAV turbo-molecular pumps and an Edwards XDS~35 scroll pump. A Gas Handling System control panel manages the mixture flow in the circuit lines, either automatically or manually. In automatic operation a CPU running a software program, that reads measures from several Pirani pressure gauges and a flow meter placed in the circuit line, manages the flow operating a number of solenoid valves. Temperatures of the experimental setup are measured by 11 thermometers of three different types (Pt1000, FeRh and RuO$_2$ resistances), connected to an AVS-45 and an AVS-47 Picowatt Resistance Bridges. The resolution of the RuO$_2$ thermometers at the lowest temperatures is 0.25~mK. The cryostat is also equipped by 4 vacuum gauges and a LHe level gauge. Inside the dilution refrigerator there are 2 capacitance gauges that measure the liquid level in the 1K Pot and in the Still. All diagnostic data are gathered, via serial and GPIB interfaces, by a PC running a LabView program which displays the readings on a synoptic window and records all the measurements.

To avoid transmission of mechanical vibrations to the bar,  the thermal links between the cold spots of the refrigerator and the experiment are kept to a minimum (n.~8 in Fig.~\ref{cryo}):
i) a couple of soft, thin OFHC copper sheets between the Mixing Chamber and the top of the suspension (n.~9 in Fig.~\ref{cryo}), and ii) 3 sheets, same as above, between the Still flange and the 0.6~K flange (n.~6 in Fig.~\ref{cryo}). These contacts assure the bar and suspension cooling when the dilution refrigerator is in operation, i.e.~below 4~K. Above this temperature gas conduction provides the heat removal by inserting a few mbar of gaseous helium in the experimental chamber. The gas is then removed before reaching the liquefaction (about 5~K).

Cooling the cryostat to 80~K and filling the LN$_2$ dewar takes about 3 days and about 1000~L of LN$_2$. 12 hours and about 800~L of LHe are sufficient to cool the system from 80~K to 4.2~K and leave about 150~L of liquid in the dewar. The cryostat consumption, once thermalized, is about 1~L/h of LN$_2$ and about 1.5~L/h of LHe, that raises to about 2~L/h when the 1K Pot is in operation.

The minimum temperature reached by the bar during the data taking was 343~mK, read by a RuO$_2$ thermometer placed at the center of one of the bar end faces.

\subsection{RAP data collection and analysis}
The piezoelectric signals were recorded by an ADC sampled at 100~kHz running under LabView control. At low temperature, due to the low specific heat, the thermometer has sufficient sensitivity to measure the increase of the bar temperature after each shot. This additional information helped in monitoring the beam intensity. It was important to read this raise immediately after the hit, as the temperature rapidly relaxes back to its equilibrium value.  For this reason, after the cooldown and during the beam measurement, we gave up multiplexing  the thermometers, so that the thermometer on the bar end face can be read out with a rate of 1~Hz.

Fig.~\ref{jump} shows the bar temperature measured during a low temperature run: it can be clearly seen that the bar  is warmed up by the beam, and after each hit it relaxes to a higher temperature.  To deal with this problem, the measurements were started at base temperature (343~mK) and taken with successively increasing temperature.

\begin{figure}[t]
\begin{center}
\includegraphics[width=4in,height=4in]{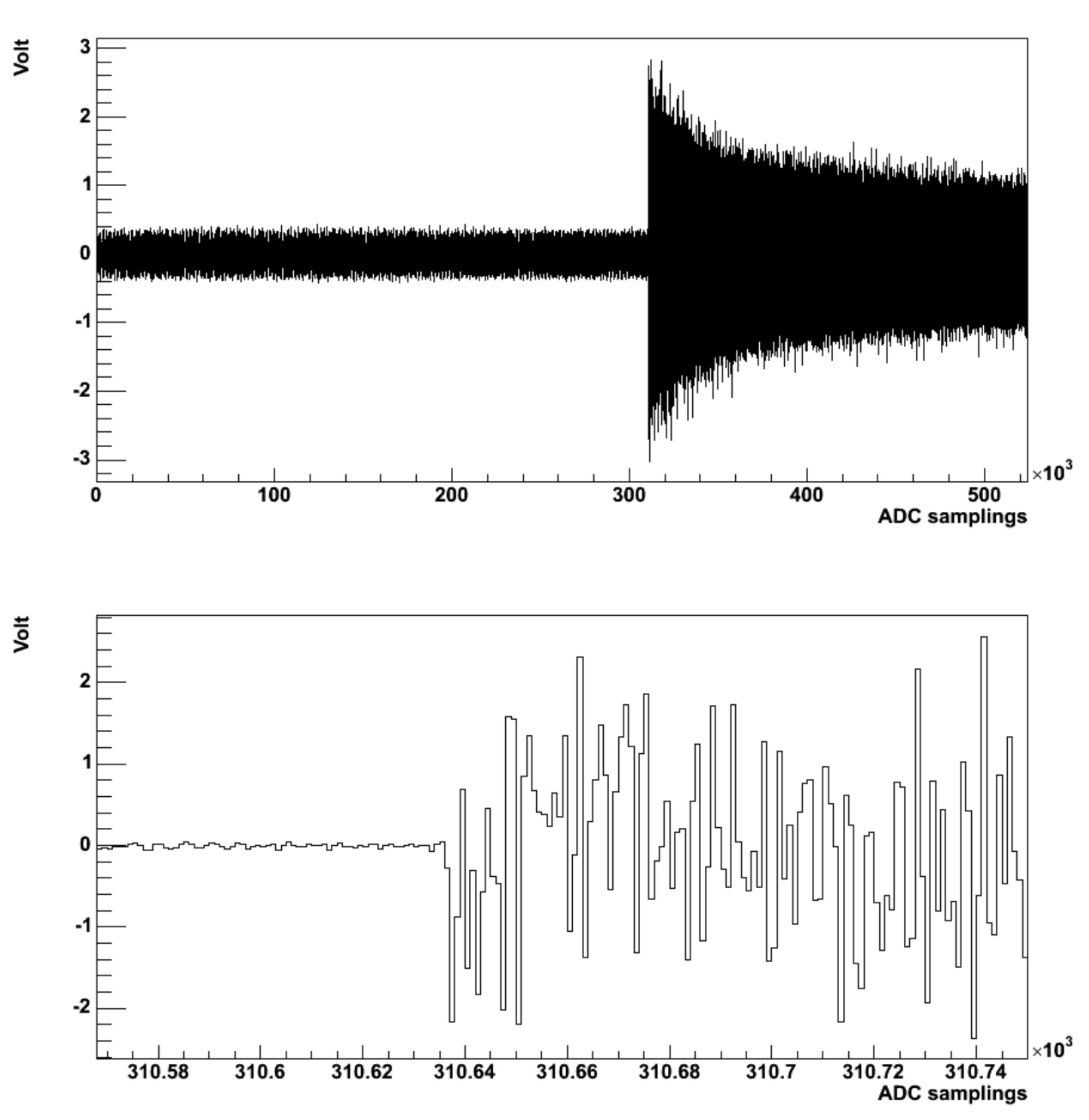}
\caption{\it Above: an example of a shot at $T$~=~0.368~K. After the shot the temperature raised to 0.405~K. The deposited energy was 30~mJ. The horizontal units are samples, the sampling time being 10~$\mu$s.  Below: the zoom of the ADC output to show the sign of the first value above the background, in this case negative.}
 \label{ADC} 
\end{center}
\end{figure}

\begin{figure}
\begin{center}
\includegraphics[width=3.8in,height=2.9in]{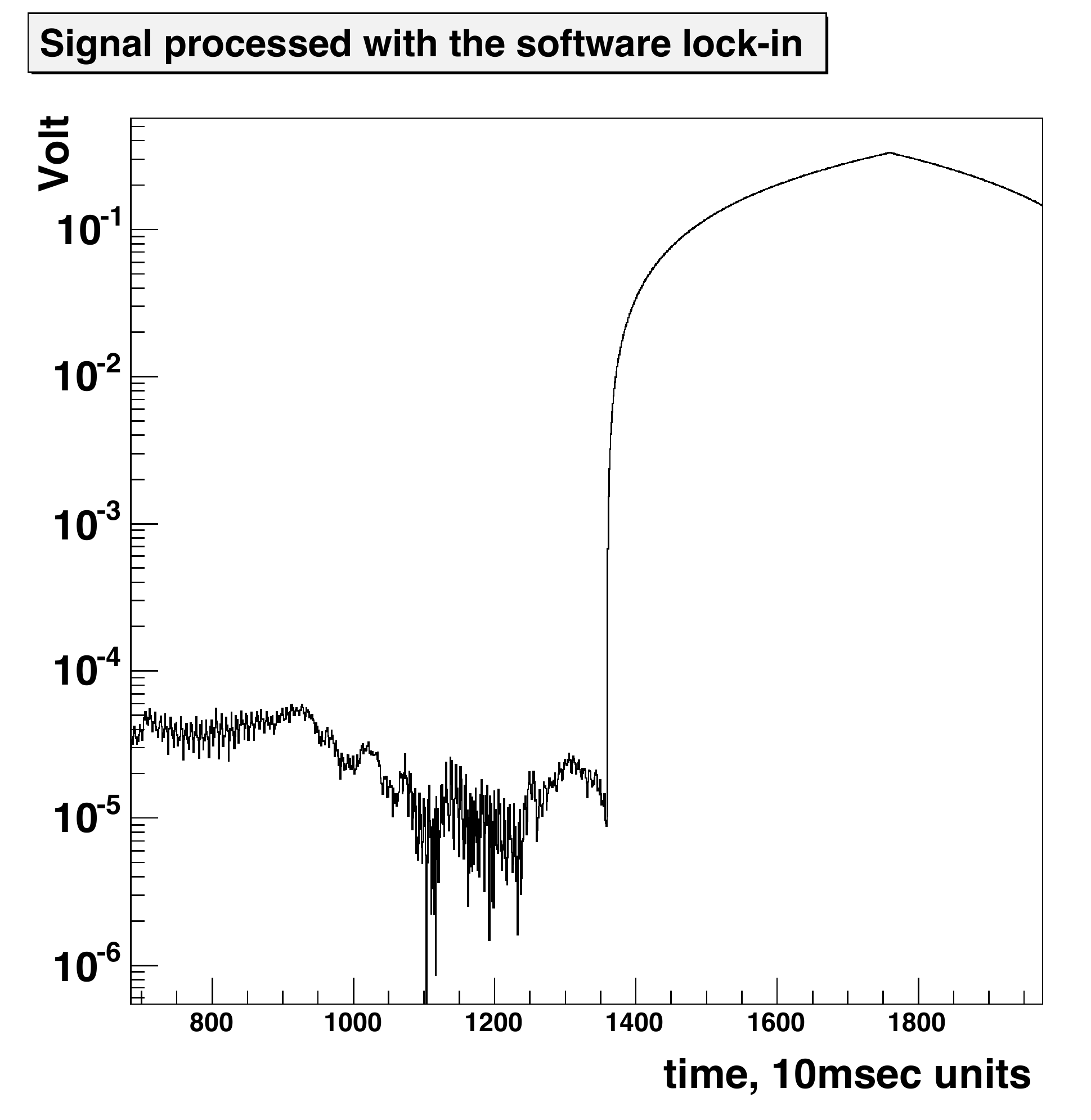}
\caption{\it The output of the filtering procedure that selects the signal component at the first longitudinal resonance $vs$ time from the start of the run. The shot is the same of  Fig.~\ref{ADC}. The filtering procedure is most useful  for small signals, when the amplitude is comparable to the noise.}
 \label{lock} 

\vskip1cm

\includegraphics[width=3.8in,height=2.9in]{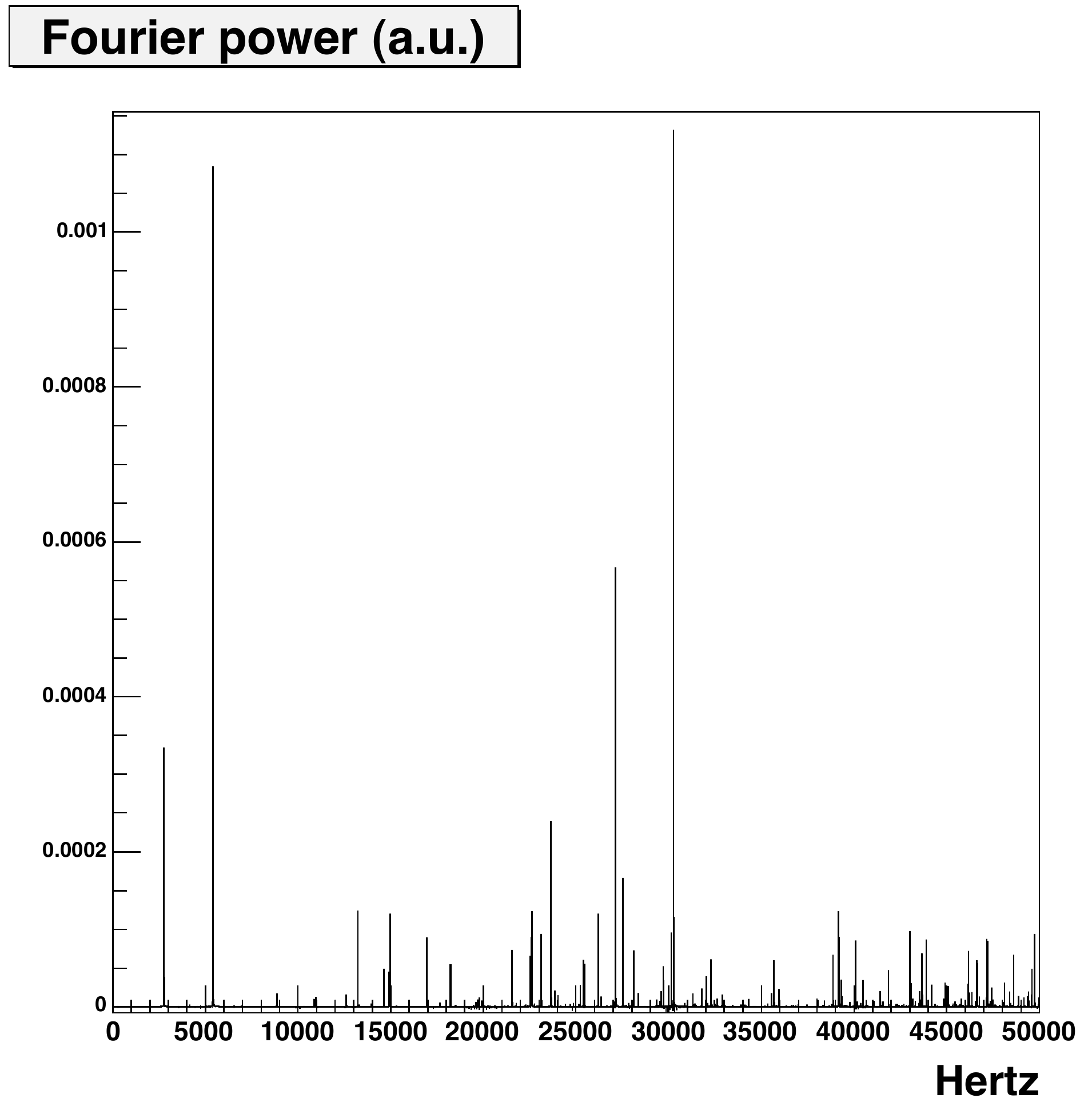}
\caption{\it The Fourier power spectrum for the average of seven signals like that shown in Fig.~\ref{ADC}, after subtracting the noise spectrum. The first two spectral  lines are the first flexural mode and the first longitudinal mode.}
 \label{Four} 
\end{center}
\end{figure}

The 100~kHz ADC data were processed  both online for a fast response and  offline with a more sophisticated procedure. Fig.~\ref{ADC} and Fig.~\ref{lock} show several important features of the signal. Fig.~\ref{ADC} shows the ADC output for a typical shot at low temperature; in the lower part of the figure the data are zoomed around the hit time in order to exhibit the sign of the first swing above the background (negative in the shown example); Fig.~\ref{lock} shows the output in volt of the filtering procedure that selects the signal at the first longitudinal resonance. The filtered output is shown $vs$ time from the start of the run.

Fig.~\ref{Four} shows the Fourier power spectrum for the average of seven signals like the one in Fig.~\ref{ADC}, after subtracting the noise spectra. The first two lines are the first flexural mode and the first longitudinal mode. We have studied in detail only the first longitudinal mode, because it is the one of interest for $gw$ detectors. However, a quick analysis of the other modes did not show any relevant difference regarding the temperature dependence.

The Fourier component at the desired frequency  $f_0$ (angular frequency $\omega_{0}$)  is extracted with a filtering algorithm known as "digital lock-in".  For a hit at time $t_0$, corresponding to sample $i_0$, we create the time series:

\begin{eqnarray*}
c_{bkg}=\sum_{i=i_{0}-N }^{i_{0}-1  }  {V_{i}\cos(\omega_{0} \frac{i}{f_{c}} )  } 
\\
c_{signal}=\sum_{i=i_{0} }^{i_{0} +N } {V_{i}\cos(\omega_{0} \frac{i}{f_{c}} )  } 
\label{lockin}
\end{eqnarray*}

\noindent where $V_i$ is the $i-th$ value of ADC output, $f_{c}$ the sampling frequency and the number of samples considered is optimized with $N$~=~400000. Similar quantities are  computed for the quadrature (sine) component, and from the two  we construct the complex amplitude before (noise) and after (signal) the hit.  Taking the difference of these two amplitudes produces the desired filtered output. A correction is applied to take into account the decay of the signal as function of the time.

\section{The RAP measurement in superconducting state}

We present in this section data of the last run of RAP at the Frascati BTF, that took place between June 30th and July 2nd  2009, just after the commissioning of the dilution refrigerator. We took data in the temperature range 0.344~--~257~K for a total of 164 beam shots on the bar and the energy deposited by each shot was in the range $1 \lesssim W \lesssim$ 70 mJ.  
The data at high  temperatures were in agreement with those of previous run and will not be discussed in this section. The 2007 measurements~\cite{rap2}, performed  at T~$\ge$~0.54~K, are analyzed in this section together with the 2009 data. Some improvements introduced in this run, like the "lock-in filter", reduced the noise and increased the sensitivity at small energies.
	
The data taken at temperatures in the range 0.9~$\le$~$T$~$\le$~2~K (i.e.~above  {$T_{c}$)  and $T$~$\le$~2~K, reported in Fig.~\ref{linea}, show a linear correlation between $B$, the maximum amplitude of the first longitudinal mode, and  the released energy $W$, in agreement with the model of section~\ref{TAM}.  The data of 2009 appear to be of  better quality than those of 2007,  due to the improvements in the beam stability and in the analysis procedure.

\begin{figure}[t]
\begin{center}
\includegraphics[width=4.1 in,height=4.1in]   {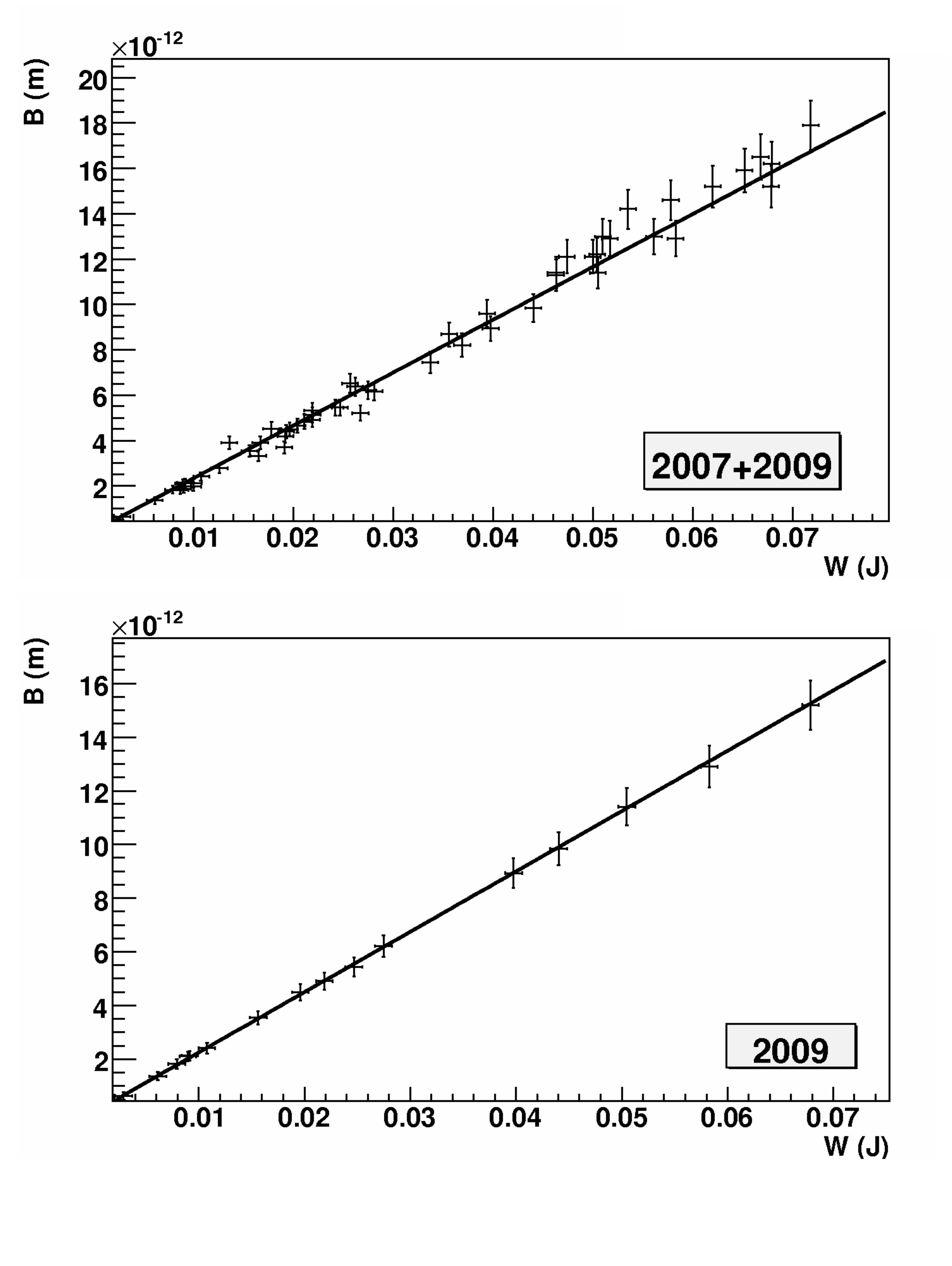}
\vskip-0.5cm
\caption{\it The measured Amplitude of the first longitudinal mode $B^{exp}$  $vs$ the energy $W$ deposited in the bar in the $n$ state  with 0.9~$\le$~T~$\le$~2~K. The solid line is a linear fit constrained to the origin. The top figure represents all data from the 2007 and 2009 runs, while the bottom one shows only the 2009 data. The result of the fits are:\newline
$p_{0}=(2.33 \pm 0.02)\times10^{-10}$  [m/J], $\chi ^2/$ndf = 51.69 / 55 (2007+2009 plot)\newline
$p_{0}=(2.24 \pm 0.05)\times10^{-10}$ [m/J], $\chi ^2/$ndf = 0.4554 / 15(2009 plot)\newline
The 2009 data have a better $\chi ^2$ due to the improvements in  the beam stability and in the data analysis.}\label{linea} 
\end{center}
\end{figure}

\begin{figure}[t]
\begin{center}
\includegraphics[width=4.3in,height=3.7in]{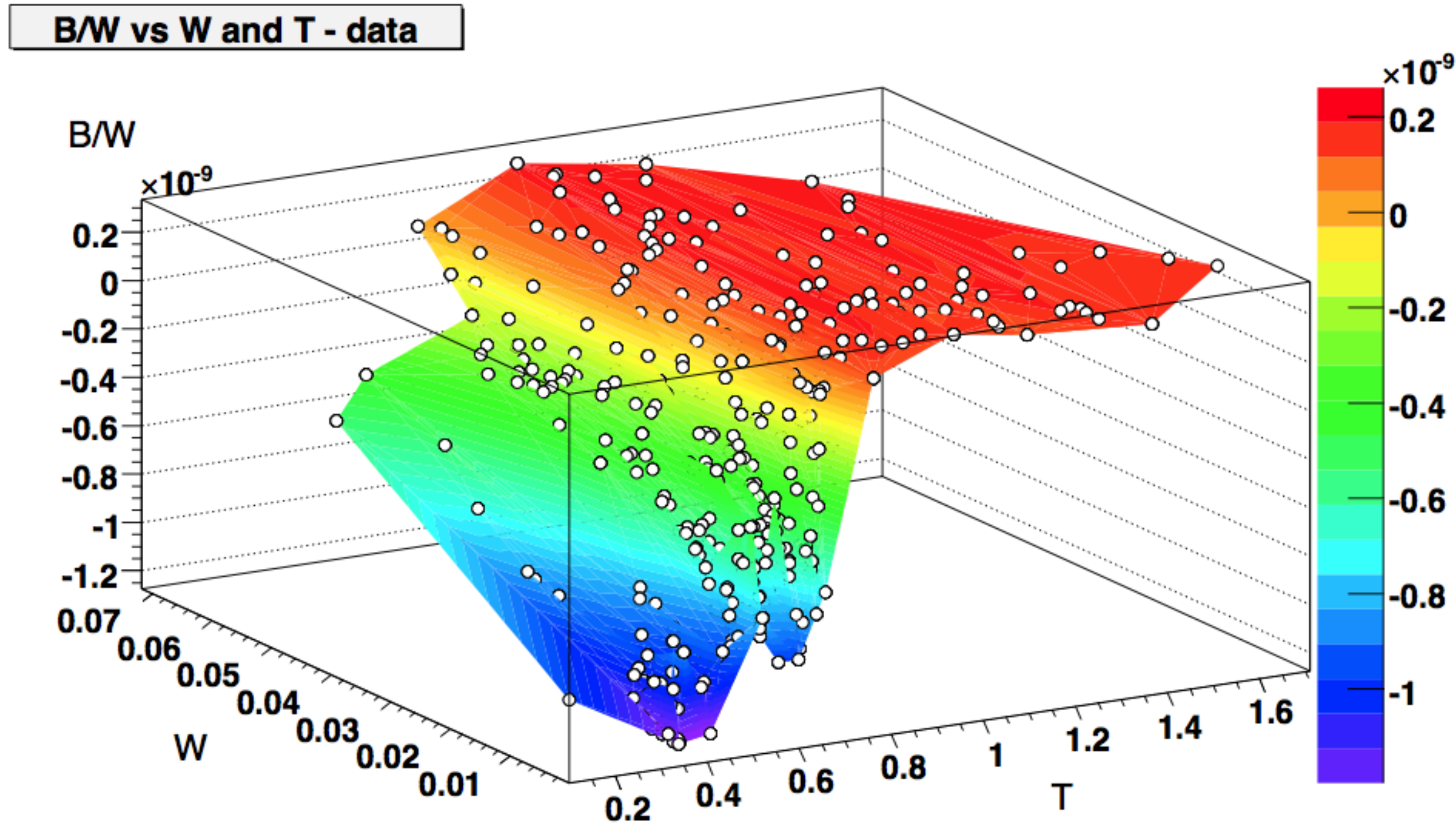}
\caption{\it Synoptic view of the data for temperature $T$~$\le$~1.6~K, the transition temperature is about 0.9~K. The plot shows the measured $B/W$ (with sign) $vs$~temperature $T$ and deposited energy $W$.  The most relevant feature of this plot are: a constant value of $B/W$ for $T$~$\ge$~$T_c$,  the change of sign of $B/W$ for $T\le$~$T_c$ and the dependence on $W$ of $B/W$ for $T$~$\le$~$T_c$. The experimental data are  the open circles. The shadowed regions are interpolations of the data. The point at  the lowest temperature $T$~=~0.14~K is obtained from the cosmic ray NAUTILUS data.}
\label{scat} 
\end{center}
\end{figure}

\begin{figure}[t]
\begin{center}
\includegraphics[width=4.5in,height=4.5in]{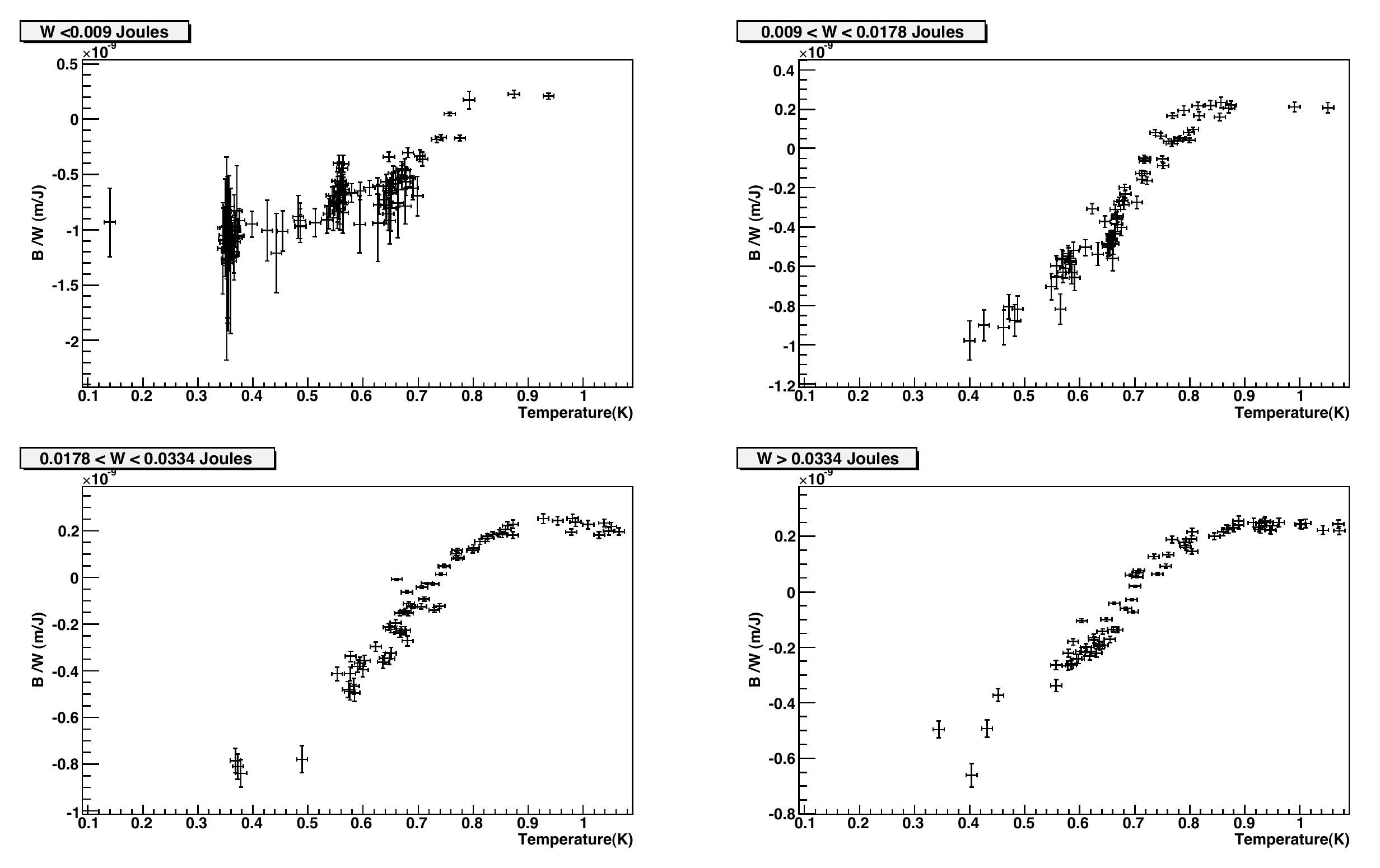}
\caption{\it Summary of the data for temperature $T$~$\le$~1.1~K. The plots show the projections of the data of Fig.~\ref{scat} in 4 energy interval.  Note that, due to the strong dependence on $W$,  data points sometimes do not overlap well.}
\label{proiet} 
\end{center}
\end{figure}

Both sets of data for temperature $T$~$\le$~1.6~K of the years 2007 and 2009 are shown in  Fig.~\ref{scat}. This plot shows  the measured $B/W$ (with sign) as function of the temperature and  of the deposited energy $W$. The most relevant features of this plot are: 
\begin{itemize}
\item a constant value of $B/W$ for $T$~$\ge$~$T_c$,
\item a change of sign of $B/W$ for $T$~$\le$~$T_c$ and
\item a nonlinear dependence of $B$ on $W$  for $T$~$\le$~$T_c$, not predicted by the model (Eqn.~(\ref{exp})).
\end{itemize}
A change of sign for superconductive aluminum is to be expected, because the effect due to the $s \to n$  transition can lead to a negative sign due to the competitive terms in Eqn.~(\ref{exp}). We recall that the sign of $B$ is inferred by the sign of the first value of the ADC over the noise after the beam shot (see Fig.~\ref{ADC}b). The sign is positive for an expansion, negative for a contraction. The plot includes the measurement of NAUTILUS with cosmic rays described in section~\ref{cosmici}. This value has been obtained from the ratio of the NAUTILUS data at $T$~=~0.14~K and the NAUTILUS data at $T$~=~2~K. The NAUTILUS point correspond to a value of the energy $W$~$\sim$~0.5~$\mu$J.
	
Fig.~\ref{proiet} shows the projection of these data in 4 intervals of deposited energy $W$.  This figure helps to understand the data behavior, but is important to note that some time data don't overlap well, due to the strong dependence on $W$. The error bars are due to the combination of  different sources: 
	
\begin{itemize}
\item the noise in the measurement of the vibration amplitude: $\pm$~1.3~$\times$~10$^{-13}$~m, 
\item the uncertainty on the deposited energy (due to a reading error  of the beam current):  $\pm$~0.8~mJ, 
\item the error in the temperature measurement, that also takes into account the local increase in the bar temperature after every shot and a possible non uniform profile of the temperature along the bar: $\pm$~0.01~K. 
\item an overall systematic error of the order of $\pm$~6\%, that accounts for the slightly different set-up and analysis procedures adopted in the 2007 and 2009 runs. 
\end{itemize}

Fig.~\ref{proiet} shows that the data at the lowest deposited energies have a simple behavior: a plateau at very low temperatures, a plateau at higher temperature (above $T_c$) and a transition region in between. The first plateau disappears in the plots for higher energies.
	
As our investigation is aimed at understanding the interactions of cosmic rays with a $gw$ detector, we need a model to make prediction of $B/W$ at very small value of $W$: we have used the model described in section~\ref{TAM},  adding to it, as suggested by the data, a possible saturation of the $s \to n$ transition effect, due the high energy density in the volume crossed by the beam.

We can estimate this saturation effect  starting from the radius of the cylindrical volume that switches  from  the $s$ to $n$ state around a particle leaving 195~MeV in the RAP aluminum bar (195~MeV is the mean value of the energy loss per each 510~MeV electron of the BTF beam).  This critical radius $R_{c}$ depends from the energy necessary to activate the transition. This energy can be computed in several ways \cite{wood}, yielding consistent results.
 Using Eqn.~(\ref{rHS}) and (\ref{swisn})  we have:

\begin{equation}
R_{c}(T)=\sqrt{\frac{W}{\pi \rho l C_I(T)}}
\end{equation}

where $l \sim 2R$ is the length of the path of the electron inside the bar.

Therefore, using the value of specific heat of Al5056 previously discussed and reported in \cite{barucci2010},  we obtain $R_{c}$~$\sim$~1~$\mu$m if $T$~=~0.5~K. The total cross surface interested by this transition for $N=10^{9}$ electrons is of the order $N \pi R_{c}^2 \sim$~30~cm$^{2}$, comparable to the beam cross section  of the BTF beam, typically $\sim$~20~cm$^2$. This is a crude estimate, but it shows that this effect can produce a non linear (with respect to the deposited energy $W$) response in the RAP data.

In order to extrapolate the RAP results to values of $W$ and $T$ outside the measured range, we used the following four parameters fit to the data for  $T< T_{c}$:

\begin{equation}
\frac{B}{W}=a+(b(T)-a) \exp \left(\frac{-W} {p_{0}~\rho ~C_{I}(T)}\right)
\label{efit}
\end{equation}

\begin{equation}
b(T)=p_{1} +p_{2} T+p_{3} T^{2}	
\end{equation}

Here $a\sim$~2.25~$\times$~10$^{-10}$ m J$^{-1}$ is the constant value of $B/W$ for $T>T_c$  obtained from Fig.~\ref{linea} and $b(T)$ the value of $B/W$ for $T<T_c$ and $W \to 0$. $b(T) \sim - 10^{-9}$ m J$^{-1}$ for $T$~=~0.5~K is  a function weakly dependent on $T$ and accounts for small variations of physical parameters  at low temperatures. $C_{I}$ is the integrated specific heat between $T$ and the critical temperature, as defined in Eqn.~(\ref{swisn}) and computed from the numerical values of Ref.~\cite{barucci2010}. Eqn.~(\ref{efit}) derives from the consideration that if an electron crosses a region that has already undergone the $s \to n$ transition, the response is the one of the $n$ state\footnote{Consider a beam bunch of high intensity, lasting a few nanoseconds (a time much shorter than the relaxation time needed to restore the $s$ state) and involving a volume $V_{0}$ of the bar: the electrons that follow will probe a variable fraction of the volume $V_{0}$   switched from the $s$ to the $n$ state. The volume $dV$ that undergoes the the $s \to n$ transition for a deposited energy $dW$ is
\begin{equation}
dV\sim  \left( 1-\frac {V}{V_{0}} \right) \frac{dW}{\rho C_{I}}
\end{equation}
}. The parameter $p_{0}$ is the  bar volume intercepted by the electron beam. 

In the measurement on a niobium bar~\cite{rap1}  the complex pattern of Fig.~\ref{scat}, with the non linear behavior in $W$, was not observed. This  is consistent with our model if we consider that niobium has a higher $T_{c}$ and specific heat is much larger: $C_{I}$ of niobium is indeed about two order of magnitude larger than $C_{I}$  of aluminum for comparable value of the integration interval $T_{c}-T$. Therefore the beam intensity in the niobium measurement was not enough to see saturation effects.

\begin{table}[htdp]
\vskip0.5cm
\begin{center}
\begin{tabular}{|cc|c|c|}
\hline
parameter& &value\\
\hline
$p_{0}$ & [m$^3$]  &$ (1.88 \pm 0.06)\times10^{-3}$ \\
$p_{1}$ & [m J$^{-1}$]  &$ (0.99 \pm 0.13)\times10^{-9}$ \\
$p_{2}$ & [m J$^{-1}$ K$^{-1}$]  &$  (-1.31 \pm 0.43)\times10^{-9}$  \\
$p_{3}$ & [m J$^{-1}$ K$^{-2}$]  &$ (-3.0 \pm0.37)\times10^{-9}$  \\  
\hline
\end{tabular}
\caption{\it Parameters for the fit of Eqn.~(\ref{efit}) to the RAP data.}
\end{center}
\label{fitr}
\end{table}

The result of  the fit of Eqn.~(\ref{efit})  with these 4 parameters are given in Table~2. 
The data fit gives a $\chi^{2}/d.o.f.$~=~368/286~=~1.29, slightly larger than one; this suggests that there are effects not taken into account.

Indeed, this simplified transition model cannot include all the effects: for example, the beam profile, the shower development inside the bar, the uncertainty of the critical temperature, possible unhomogeneity of the material, etc.~are not accounted for. It is however remarkable that a very simple expression can describe the complicated pattern of Fig.~\ref{scat}.

\begin{figure}[t]
\begin{center}
\includegraphics[width=4in,height=3in]{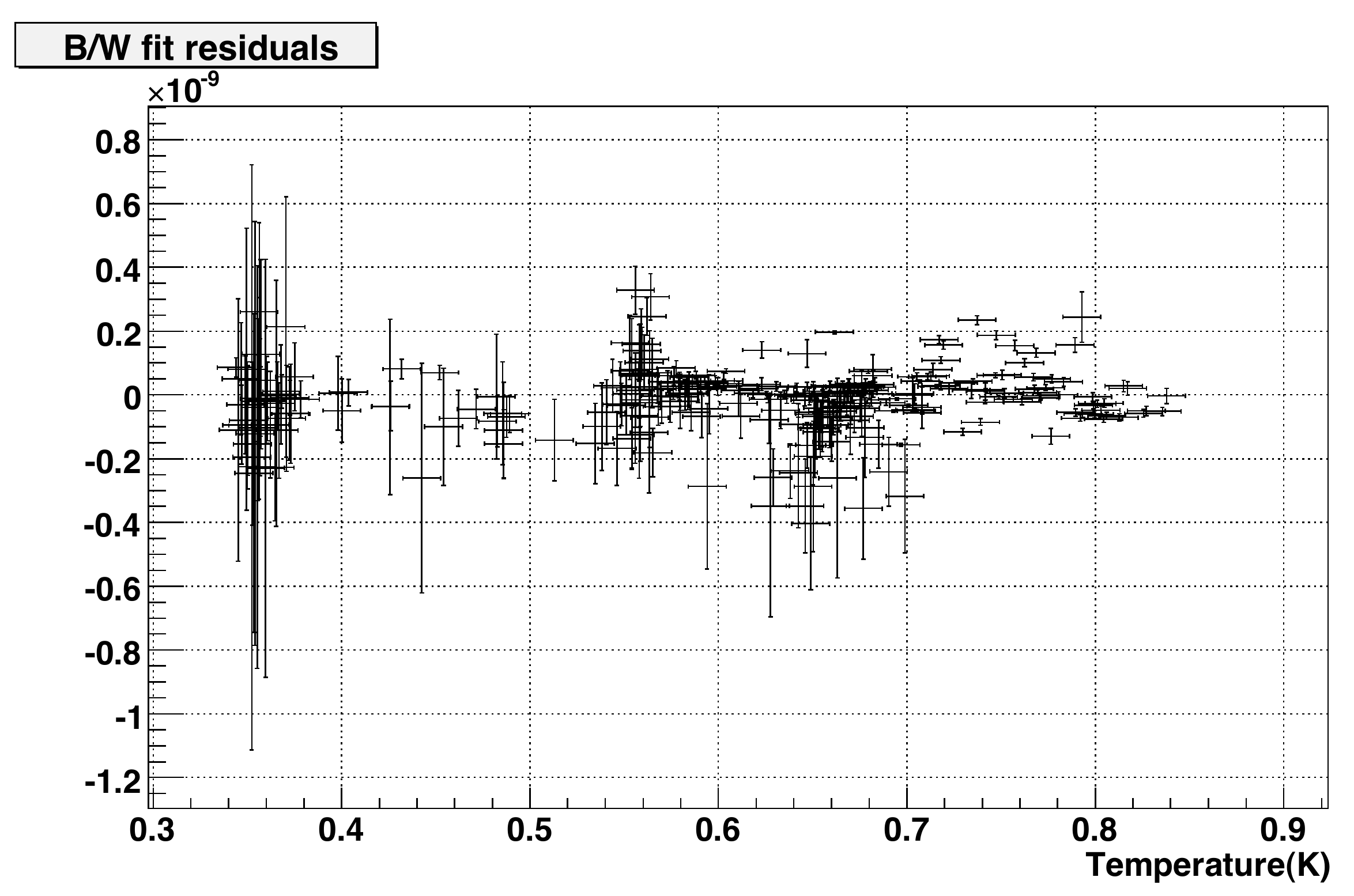}
\caption{\it Residuals of the fit of Eqn.~(\ref{efit}) to the RAP data $vs$ temperature, overlapping on all energies.}
\label{resid} 
\end{center}
\end{figure}

\begin{table}[htdp]
\begin{center}
\begin{tabular}{|c|c|c|}
\hline
Temperature & Experimental $B/W$ & Predicted $B/W$  \\
(K)& [10$^{-10}$ m J$^{-1}$] &[10$^{-10}$ m J$^{-1}$] \\
\hline
264          & 2.14~$\pm$~0.13 & 2.23       \\
71           &  2.18~$\pm$~0.13 & 2.23       \\
4.5          &  2.09~$\pm$~0.13 & 1.80       \\
1.5          &  2.25~$\pm$~0.13 &  1.72          \\   
0.8          & --0.95~$\pm$~2.4 & --7.3  \\
0.7          & --4.2~$\pm$~2.5 & --7.9  \\
0.6          & --6.9~$\pm$~2.5 & --7.7  \\
0.35         & --10.8~$\pm$~2.3 & --9.1      \\
\hline
0.14         & --11.1~$\pm$~1.8 & n/a      \\
\hline
\end{tabular}
\caption{\it Summary of the values of $B/W$ derived from measurements on Al5056. For $T>T_{c}$ the table shows measured data with errors due to the 6\% systematic error. Both values and errors for $T\le T_c{}$ are {obtained by the fit in the low energy limit} $W\to$~0, taking into account the error matrix. The last value at $T$~=~0.14~K was out of the RAP experimental reach, and is derived by extrapolating the fit of Eqn.~(\ref{efit}) also with respect to temperature. The right column shows the predictions of the model (see section~\ref{TAM}).}
\end{center}
\label{summa}
\end{table}

Fig.~\ref{resid} shows the fit residuals (measured data minus the fit predictions) $vs$ temperature.  The energy dependence seems to be well {reproduced} with the exception of the region just below $T_{c}$ where there are more scattered points.

Table~3 summarizes the RAP results of $B/W$ for Al5056  obtained  with the RAP data from room temperature down to 0.14~K. The value and the errors  for $T < T_{c}$ are obtained from the fit of Eqn.~(\ref{efit}) for $W\to$~0. The last value was obtained by extrapolating also with respect to the temperature down to $T$~=~0.14~K,  a temperature relevant for the comparison with NAUTILUS data that was, unfortunately, beyond the reach of our refrigerator. We note that the model described in section~\ref{TAM} is quite accurate only for ${T  > T_{c}}$. For ${T  < T_{c}}$ there are the discrepancies that could be due either to a failure of  the model or  to uncertainties in the Al5056 superconductive parameters.

Using this table we can estimate:
\begin{equation}
\frac {B/W_{T=0.14\rm{K}}}{B/W_{T=1.5 \rm{K}}}=4.9 \pm 0.8
\end{equation}
in good agreement with the value 4.3~$\pm$~1.5 obtained comparing cosmic rays detected in  NAUTILUS at $T$~=~0.14~K and $T$~=~3~K. We assume in the next section that this value doesn't change in the range 1.5~$\div$~4.5 K.

\section {Cosmic Rays in $gw$ acoustic detectors - interpretation in the light of RAP results}\label{cosmici}

The aim of this paper is to use the RAP results to interpret the cosmic ray signals detected in the EXPLORER and NAUTILUS $gw$ antennas. To this purpose,  we first summarize  the most relevant results on this issue \cite{Ronga:2009zza, Astone:2008xa}.

The ultra-cryogenic acoustic $gw$ detector NAUTILUS \cite{rog,Astone1991} is operating since 1996 at the INFN Frascati Laboratory, at about 200 meters above sea level: it consists of a 3~m,  2300~kg, Al5056 alloy bar. We consider here the run of 1998, when NAUTILUS was operated at 140~mK. The quantity to be observed  in this kind of detector (the "$gw$ antenna output") is the vibrational amplitude of its first longitudinal mode of oscillation. This is converted by means of an electromechanical resonant transducer into an electrical signal which is amplified by a dc-SQUID superconducting amplifier. The bar and the resonant transducer form a coupled oscillator system, with two resonant modes, whose frequencies were, in that run, $f_-$~=~906.40~Hz and $f_+$~=~921.95~Hz. NAUTILUS  is equipped with a cosmic ray detection telescope made of seven layers of gas detectors (streamer tubes) for a total of 116 counters \cite{coccia}.
 
The $gw$ detector EXPLORER \cite{expl1,longterm}, similar to NAUTILUS, was located in CERN (Geneva-CH) at about 430 meters above the sea level.  Scintillator counters were installed at EXPLORER in 2002, using scrap equipment recovered after the LEP shutdown. The two detectors have a long record of coincidence runs~\cite{ref1}, also with other detectors~\cite{ref2}, to search for $gw$ signals.

The signal expected in a $gw$ detector like NAUTILUS, as a consequence of the interaction of a particle releasing an energy $W$ is \cite{cosmico1, cosmico2, cosmico3}, according to the model described in section~\ref{TAM} is ($W$ in GeV units):

\begin{equation}
E\sim \frac{7.64}{2}\times10^{-9}~W^2~\delta^2 \hskip1cm  [\rm{K}]
\label{ww}
\end{equation}

\noindent where the bar oscillation energy $E$ is expressed, as usual in the antenna jargon, in Kelvin units (1~K~=~$1.38 \times 10^{-23}$~J),  the numerical constant  is the value computed using the thermal expansion coefficient and the specific heat of pure aluminum at 4~K and $\delta$ is a parameter that describes the difference respect to pure aluminum at 4~K.
In the previous section we have shown that RAP has measured $\delta_{n}$~=~1.16  above the $s$ transition temperature or $\delta_{s}$~=~5.7~(=~4.9~$\times$~1.16) for superconductive Al5056. The vibrational energy $E$ of the first longitudinal mode of oscillation  is proportional to the square of the Amplitude (in an oversimplified model, $E= \frac{1}{4}M \omega_0^2 B^2$). The constant 7.64~$\times$~10$^{-9}$ applies if the energy is released in the bar center. If the energy is uniformly distributed along the bar, as in the case of extensive air showers (EAS),  this value is reduced by a factor 2. 

Under simplified approximations on the development of the electromagnetic shower in the bar, we can derive \cite{cosmico1,cosmico2,cosmico3}  the relation between the vibrational energy detected in the bar and the density $\Lambda$ of secondaries in the shower:
\begin{equation}
E=4.7 \times 10^{-10}~\Lambda^2~ \delta^2  \hskip1cm   [\rm{K}]
\label{theo}
\end{equation}

The plot of vibrational energy $E_{exp}$ $vs$ particle density $ \Lambda$ is shown in Fig.~\ref{resp}. In this figure we show the events detected by NAUTILUS, both at 140~mK and at 2.6~K, as well as by EXPLORER at a temperature of about 3~K. We clearly see a difference of more than one order of magnitude between the measurements taken with  aluminum in the $s$ state and those in $n$ conduction state. From this plot we can estimate a value of $\delta_s=5.0 \pm 1.8$ in good agreement with the value derived by the RAP experiment: $\delta_s=5.7 \pm 0.9$.

\begin{figure}[t]
\begin{center}
\vskip-1cm
\includegraphics[width=5in,height=4in]{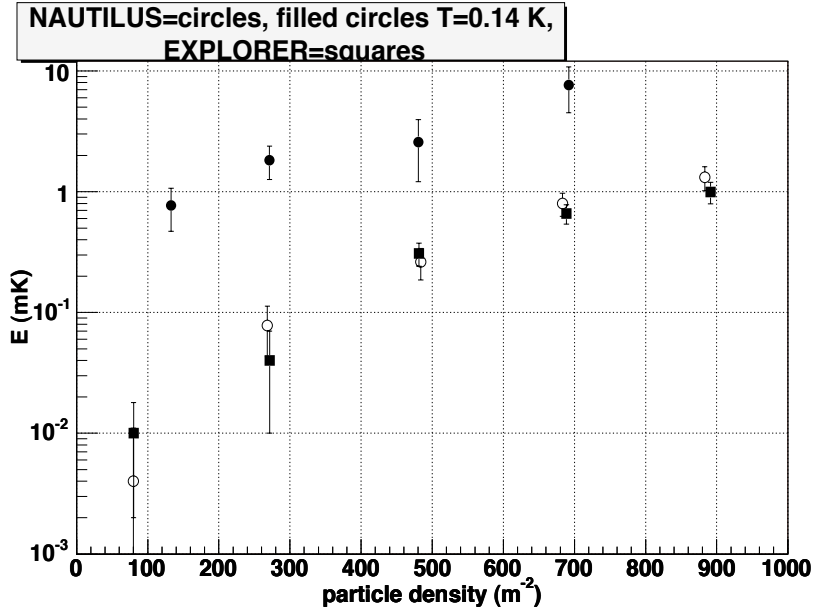}
\caption{\it Averages of signals with energy $E_{exp}\le0.1$~K, grouping data in ranges of particle density $\Lambda$. Filled circles NAUTILUS  at $T$~=~0.14~K, open circles  NAUTILUS at $T$~=~3~K, filled squares EXPLORER at $T$~=~3~K. The data gathered at $T$~=~0.14~K are roughly one order of magnitude larger than those collected at $T$~=~3~K. From Ref.~\cite{Astone:2008xa}.
\label{resp} }
\end{center}
\end{figure}

We now briefly discuss the event rate of cosmic rays in $gw$ detectors.  The cosmic ray event rate in NAUTILUS and EXPLORER has been evaluated in the past considering three different event categories: i) pure electromagnetic showers, responsible for most of  the high energy events detected in the bar detectors; ii) showers produced by muons and iii) showers produced by hadrons in the bar. We use Eqn.~(\ref{ww})  with the correction $\delta_n=1.16$ for the response of an Al5056 bar in the $n$ state. The rate of electromagnetic air showers (EAS) is computed starting from the empirical relation due to G. Cocconi~\cite{cocco}. The event rate due to muon and hadron interactions inside the bar was computed using the GEANT package~\cite{geant}, to simulate the antenna  and the CORSIKA Monte Carlo~\cite{heck}, as input to GEANT, to simulate the effect of the hadrons produced by the cosmic ray interactions in the atmosphere, assuming a cosmic ray "light" composition. The Monte Carlo simulation represents 1 year of data taking.

The results are  shown in Table~\ref{table_1}. The energy in the first longitudinal mode $E$ (first column) is proportional to the square of the absorbed energy $W$.  

\noindent The rate of the the events scales as  $W^{-0.9}$. This is because the cosmic ray integral spectrum is well described by a power law  $E_{CR}^{-\beta}$ with  $\beta\sim1.7$  for cosmic ray primaries up to the so called "knee" at $E_{CR}=10^{15}$~eV and $\beta\sim2$ at higher energies. 
     
\begin{table}[h]
\vskip0.5cm
\centering
\begin{tabular}{|c|c|c|c|c|c|c|c|}
\hline
Vibrational $E$&Deposited $W$&Total\\
$[\rm K]$&[GeV]&[events/day]\\
\hline 
$\ge10^{-5}$&$\ge44.5$& 107  \\
$\ge10^{-4}$&$\ge141$& 14.5   \\
$\ge10^{-3}$&$\ge445$& 1.6   \\
$\ge10^{-2}$&$\ge1410$& 0.19   \\
$\ge10^{-1}$&$\ge4450$&  0.03  \\
\hline
\end{tabular}
\caption{\it Estimated rate of antenna excitations due to cosmic rays in NAUTILUS $vs$~the vibrational energy of the longitudinal fundamental mode that such events can produce.  The value at $E$~=~0.1~K is obtained extrapolating from the lower energy values. The values in the second column are the values of cosmic ray energy that the bar needs to absorb in order to have an excitation energy $E$. Vibrational and Deposited energy are correlated by Eqn.~(\ref{ww}), under the assumption of uniformly distributed energy.}
\label{table_1}
\vskip0.5cm
\end{table}

In Ref.~\cite{cosmico2},  very large NAUTILUS signals at a rate much greater than expected were reported. In the light of the analysis reported above, it is now clear that the value $\delta_s=5.7$ must be used in Eqn.~(\ref{theo}) to compute the expected response and the expected rate. The NAUTILUS 1998 data event rate per day after the unfolding of  the background, with the procedure described in \cite{Astone:2008xa} is shown in Fig.~\ref{naut98a}. The continuos line is the prediction of Table~\ref{table_1} with $\delta_s=5.7$. We find now a good agreement between measurements and predictions; previously hypothesized exotic explanations, based on anomalous component of cosmic rays or anomalous interactions of cosmic rays with a superconductive bar can now be excluded.

Cosmic ray showers are a very important tool to verify the sensitivity of $gw$ bar detectors to signals distributed along the bar, signals similar to $gw$ even if the excitation mechanism is different. Moreover the cumulative analysis of Fig.~\ref{resp} and the agreement with the RAP measurements show that for this kind of search the sensitivity of bar detectors can be studied down to 10~$\mu$K corresponding to $B < 10^{-19}$~m.  This results is quite important for other "cumulative" analysis like the search of signals from gamma ray bursts.
\begin{figure}[h]
\begin{center}
\includegraphics[width=4.in,height=3.2in]{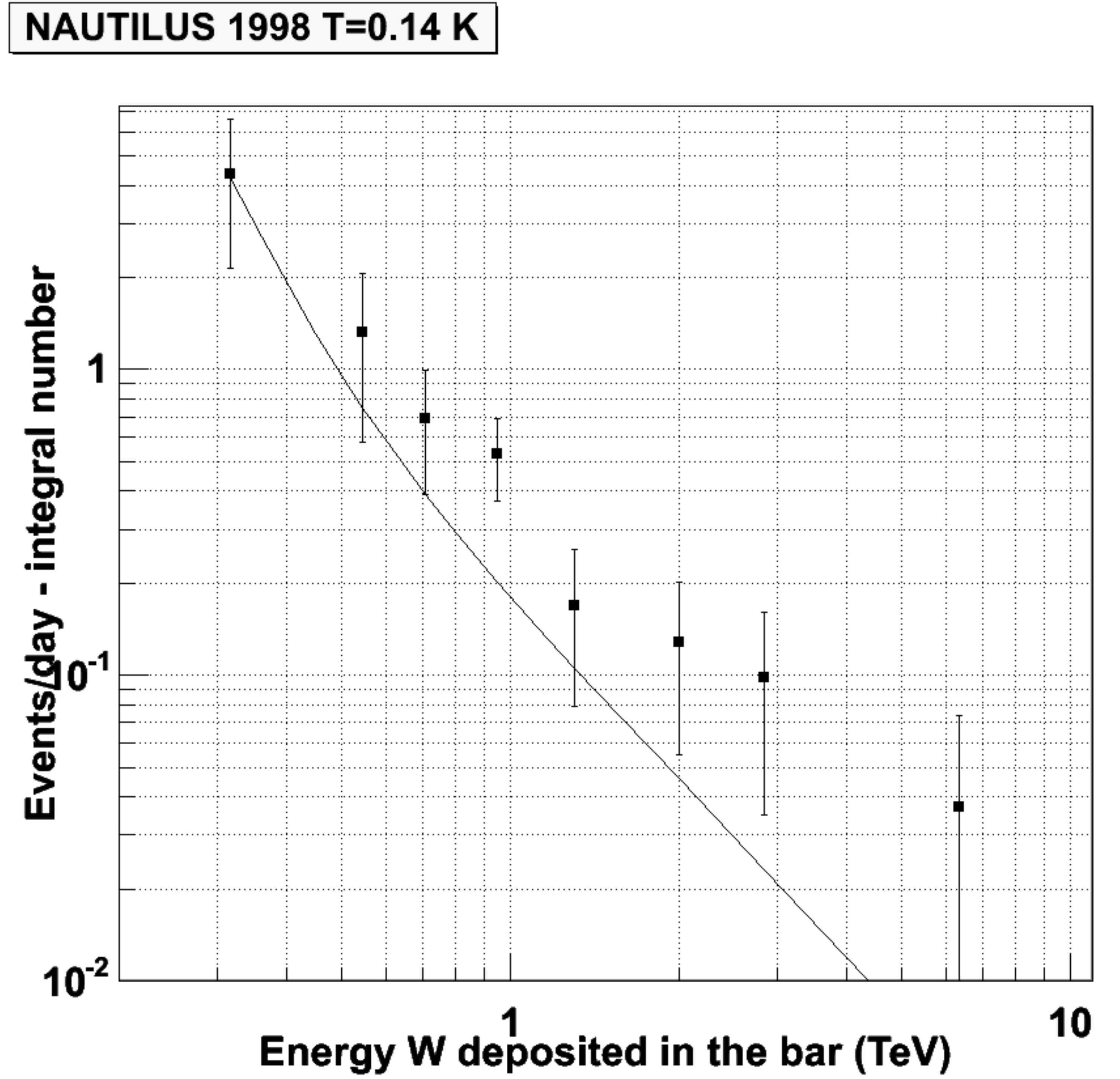}
\caption{\it NAUTILUS 1998, at $T$~=~0.14~K. The integral distribution of the event rate after the background unfolding,  compared with the expected distribution (continuous line). The prediction is computed using the data of Table~\ref{table_1} and using the value $\delta_s=5.7$ measured by RAP. The good agreement suggests the absence of anomalous components of cosmic rays or anomalous interactions of cosmic rays with a superconductive bar. Modified from Ref.~\cite{Astone:2008xa}.}
\label{naut98a}
\end{center}
\end{figure}

Finally, we remark that acoustic $gw$ detectors have no limitations due to saturation effects in detecting large signals. Indeed the largest event detected up to now has a vibrational energy in the first longitudinal mode  $E \sim670$ K corresponding to $\sim360$ TeV in the bar. The event occurred in EXPLORER on Nov 10 2006 9:40 UT.

\section {Conclusions}

We have shown that the Thermo-Acoustic Model reasonably well describes the response of a bar to the passage of ionizing particles. At high temperatures, in a normal conduction state, the prediction, only based on the knowledge of the Gr\"{u}neisen parameter, is in a good agreement with the data. In a superconductive material, the transitions between the $s$ and $n$ state complicate this picture. Due to the poor knowledge of the low temperatures parameter and to the approximations of the model, a direct measurement with a particle beam was needed to directly measure the response to ionizing particles. The RAP experiment addressed this issue. At high energy densities of the impinging particles, we have detected and studied with RAP non linear effects that complicate the data analysis.  

We have shown  that the unexpected large events detected in 1998 with NAUTILUS at $T$~=~0.14~K were due to its superconductive state. Using the RAP measurements we have regauged the rate of cosmic rays detected by the NAUTILUS and EXPLORER antennas, and shown that they are in agreement with the predictions.
  
Currently the background due to cosmic rays in acoustic bar detectors is negligible. This is because the typical sensitivity is $E$~=~1~mK;  the standard event selection requires a threshold of about 25~mK. With such a threshold we have a few events per week due to cosmic rays. These events are however very useful as a tool to continuously monitor and calibrate the acoustic $gw$ detectors.  Moreover, in the standard NAUTILUS data analysis, these events are currently vetoed and removed by the official $gw$ event list.

Cosmic rays will represent an important source of background in future higher sensitivity~\cite{Yamamoto:2008fs}, possibly superconducting, detectors, and this noise should be taken into account in both acoustic~\cite{Bonaldi:2006nj,deWaard:2003ug,coccia98} and interferometric~\cite{Punturo:2010zz} detectors. To remove this background, moving to an underground site could be necessary.

\begin{center} 
*** 
\end{center}

We gladly acknowledge precious help from our technicians M. Iannarelli, E. Turri, F. Campolungo, R. Lenci and F. Tabacchioni. This work was partially funded by the EU Project ILIAS (RII3-CT- 2004-506222).

\bibliographystyle{model2-names}
\bibliography{<your-bib-database>}



\end{document}